\documentclass[fleqn,usenatbib]{mnras}

\usepackage{newtxtext,newtxmath}

\usepackage[T1]{fontenc}

\DeclareRobustCommand{\VAN}[3]{#2}
\let\VANthebibliography\thebibliography
\def\thebibliography{\DeclareRobustCommand{\VAN}[3]{##3}\VANthebibliography}

\usepackage{graphicx}	
\usepackage{amsmath}	
\usepackage{physics}	

\newcommand{\dO}{\mathrm{d}\Omega}
\newcommand{\Oeq}{\Omega_\mathrm{eq}}

\defcitealias{ZDK21}{Paper \rm{I}}

\newcommand{\BZ}[1]{{{#1}}}

\title[Magnetic evolution of the K2~dwarf V471~Tau]{Magnetic field evolution of the K2~dwarf V471~Tau}

\author[B.~Zaire et. al]{
B.~Zaire$^{1}$\thanks{E-mail: bonnie.zaire@irap.omp.eu},
J.-F.~Donati$^{1}$,
and  B.~Klein$^{2}$
\\
$^{1}$IRAP, Université de Toulouse, CNRS / UMR 5277, CNES, UPS, 14 avenue E. Belin, Toulouse, F-31400 France\\
$^{2}$ Sub-department of Astrophysics, Department of Physics, University of Oxford, Oxford, OX1 3RH, UK}

\date{Accepted XXX. Received YYY; in original form ZZZ}

\pubyear{2022}

\begin{document}
\label{firstpage}
\pagerange{\pageref{firstpage}--\pageref{lastpage}}
\maketitle

\begin{abstract}
Observations of the eclipsing binary system V471 Tau show that the time of the primary eclipses varies in an apparent periodic way. With growing evidence that the magnetically active K2 dwarf component might be responsible for driving the eclipse timing variations (ETVs), it is necessary to monitor the star throughout the predicted $\sim35$\;yr activity cycle that putatively fuels the observed ETVs. We contribute to this goal with this paper by analysing spectropolarimetric data obtained with ESPaDOnS at the Canada-France-Hawaii Telescope in December 2014 and January 2015. Using Zeeman-Doppler Imaging, we reconstruct the distribution of brightness inhomogeneities and large-scale magnetic field at the surface of the K2~dwarf. Compared to previous tomographic reconstructions of the star carried out with the same code, we probe a new phase of the ETVs cycle, offering new constraints for future works exploring whether a magnetic mechanism operating in \BZ{the} K2~dwarf star is indeed able to induce the observed ETVs of V471~Tau. 
\end{abstract}

\begin{keywords}
Magnetic fields --  stars: magnetic field --  stars: imaging -- stars: individual: V471 Tau --  binaries: eclipsing -- techniques: polarimetric
\end{keywords}



\section{Introduction}
Several \BZ{eclipsing} binary systems display periodic eclipse timing variations (ETVs) when considering a linear ephemeris to predict the time of mid-eclipse \citep{LRR98,LR99,ZS13,BMP16}. It is estimated that around 90 per cent of the post-common-envelope binary (PCEB) systems display ETVs \citep{ZS13}. The main explanations that have been proposed to account for the existence of ETVs are associated with the presence of circumbinary bodies perturbing the orbit of the system \citep{I52} or magnetically-induced gravitational modulations caused by an active star in the system \citep{A87,A92,LRR98,L05,L06,L20,VSP16,VSB18}. In most cases, ETVs are attributed to circumbinary planet/sub-stellar components that, given their mass and orbital distance, can explain the periodicity and amplitude of ETVs \citep{PMC10,RDL13,CPS14,MPB14,HBF19,MHM20,PCC21}. However, recent investigations showed that caution must be taken when interpreting ETVs as caused by circumbinary objects \citep[e.g.,][]{M18}. In particular, some of the circumbinary objects inferred from the ETVs have been refuted afterwards using dynamical stability analysis \citep{HMW11,HHW12,HWH14,WHM12,M18,MM21} or high-resolution direct imaging of the systems \citep[e.g. V471~Tau,][]{HSP15}.

V471~Tau is a close binary system consisting of a K2~dwarf main-sequence star and a hot white dwarf \citep{NY70}. The system has a short orbital period of $P_\mathrm{orb} = 0.5211833875$\,day \citep{VWV15} and  due to tides the K2~dwarf is forced to rotate nearly synchronously with the orbital period ($P_\mathrm{rot} \approx P_\mathrm{orb}$). As in most PCEBs, cyclic ETVs are observed in V471~Tau with typical modulations of semi-amplitude $\Delta P/P_\mathrm{orb}~\approx~8.5\times10^{-7}$ (where $\Delta P$ is the difference between the observed orbital period minus the mean orbital period $P_\mathrm{orb}$) and periodicity of 30–35~yr \citep{KH11,VWV15,MGE18,L20}. \citet{GR01} analysed whether the gravity influence of a hypothetical third body could lead to the ETVs of the system. The authors found that V471 Tau would need a brown dwarf component with a mass of $\approx 0.0393 \pm 0.0038$\,M$_\odot$ and a semi-major axis of $11.2 \pm 0.4$\,AU to reconcile the amplitude and periodicity of the ETV cycle. However, an image of V471~Tau obtained with SPHERE at the Very Large Telescope (VLT) refuted the existence of the brown dwarf \citep{HSP15}. This view is supported by \citet{VCD17}, who dismissed the brown dwarf component using different arguments based on the lack of temporal variations of the rotational period of the white dwarf (that otherwise should vary with the same periodicity of the ETVs due to the barycenter wobbling). 

Alternative effects of magnetic origin have thus been put forward as the most probable cause of ETVs in V471~Tau \citep[e.g.,][]{A92,VSP16,NSZ18,NSK20,L20}. Despite differences between the proposed models, a common feature that they all share relies on the magnetism of the active component in V471 Tau – i.e., the K2~dwarf star. The Applegate effect \citep{A92} explains ETVs as an indirect outcome of the redistribution of angular momentum within the convective zone of the K2~dwarf throughout a magnetic cycle. The main idea behind the model is that the redistribution of angular momentum  causes temporal modulation of the gravitational quadrupole moment of the K2~dwarf. This increases (resp. decreases) the gravitational field at the orbital plane forcing the white dwarf component to orbit closer to (resp. further from) the K2~dwarf and with shorter (resp. longer) periods to conserve the total angular momentum of the system (thus creating ETVs). Besides, activity studies suggest a cyclic nature for the magnetism of the K2~dwarf yielding a putative period of about 13\,yr \citep{IET05,KRJ07,PS08,KKO21}. However, the feasibility of the Applegate mechanism in V471~Tau has been debated ever since \citet{A92} as it requires significant variations of the differential rotation that are yet to be detected at the surface of the K2~dwarf \citep[see discussions of][]{L05,L06,VSP16,VSB18,ZDK21}.

\citet{L20} proposed a new mechanism (hereinafter the Lanza mechanism) that requires lower variations of the differential rotation at the surface of the K2~dwarf to explain the ETVs. This new model is based on the existence of a non-axisymmetric gravitational quadrupole moment induced by a non-axisymmetric stationary field throughout the convective zone of the K2 dwarf. Similar to the Applegate effect, the idea behind the Lanza effect is that the modulation of the gravitational field along the line joining both stars generates ETVs. However, the Lanza effect provides a novel approach to the source of variation of the gravitational field, which results from a non-axisymmetric stationary magnetic field that is forced to librate around the Lagrange L1 point of the system or to circulate monotonically in the orbital plane. The \citeauthor{L20} mechanism has been shown to reduce by at least an order of magnitude the required fluctuation amplitude of the differential rotation with respect to the Applegate effect. Nevertheless, in order for the Lanza effect to explain the ETVs of V471~Tau the non-axisymmetric field needs to librate/circulate with a period of $70$\,yr, which disagrees with the 13\,yr activity cycle proposed \BZ{from recent} observations of the K2~dwarf \citep{KKO21}. Therefore, the origin of ETVs on V471~Tau is still unclear and demonstrating whether an Applegate effect, a Lanza effect, or another effect of magnetic origin  operates in the system requires dedicated studies of the K2~dwarf magnetism. 

Recently, \citet[][hereafter \citetalias{ZDK21}]{ZDK21} reported first large scale surface magnetic maps and offered new differential rotation measurements of the K2~dwarf for two different epochs (November/December 2004 and December 2005). They found that the K2~dwarf exhibits significant fluctuations in its differential rotation amplitude (ranging from the solar value to about twice the solar differential rotation in a year) and it is not always rotating as a solid body as it was reported to in an early study \citep{HAS06}. Despite providing useful information to disentangle the magnetic effects proposed to explain ETVs in V471~Tau, this initial study only probed a maximum of the ETVs cycle at which differential rotation is not expected to peak in the Applegate scenario. Additional surface maps and shear measurements probing different phases of the ETV cycle are thus still needed to determine the fluctuation amplitude of the surface shear and to search for a possible long term evolution, perhaps following the prediction of  \citeauthor{A92} or \citeauthor{L20}, of the surface magnetic field of the K2~dwarf. 

In this study, we reconstruct new large-scale magnetic field maps and perform new differential rotation measurements of the K2~dwarf of V471~Tau in December 2014/January 2015, probing a new phase of the ETV modulation cycle in which the observed orbital period is close to the mean orbital period $P_\mathrm{orb} = 0.5211833875$\,day \citep{VWV15}. Section~\ref{sec:obs} describes the spectropolarimetric observations and, Section~\ref{sec:results}, presents the tomographic reconstructions and the differential rotation measurements. Finally, we discuss our results and conclude in Section~\ref{sec:conclusions}. 


\section{Observations}\label{sec:obs}

We use spectropolarimetric observations of V471~Tau collected with \verb'ESPaDOnS'  at the Canada-France-Hawaii Telescope. The optical spectropolarimeter \verb'ESPaDOnS' covers wavelengths from $370$ to $1,000$\,nm  at a resolving power of $65,000$ \citep{D03,DCL06}. Our data set consists of 236 unpolarised (Stokes $I$), and 59 circularly polarised (Stokes $V$) profiles acquired in 11 nights spread between 20 December 2014 and 12 January 2015. Circularly-polarised spectra are computed combining 4 sub-exposures of 200\,s each taken at different orientations of the polarimeter retarders combined in an optimal way to minimize potential spurious signatures and to remove systematics in the circularly polarised spectra \citep{DSC97}. The data reduction was carried out with the pipeline \verb'Libre-ESpRIT' optimized for \verb'ESPaDOnS' observations \citep{DSC97}. The observational logbook is given in Table~\ref{tab:Data}. Circularly polarised spectra show peak signal-to-noise ratios (SNRs) ranging from 122 to 212  (per 1.8\,km\,s$^{-1}$ spectral pixel), with a median of 184. Orbital cycles $E$ are computed according to the ephemeris of \citet{VWV15}:
 \begin{equation} \label{eq:ephemeris}
 \mathrm{HJED} \qq{=}  2445821.898291 + 0.5211833875\times E,
\end{equation}
where phase 0.5 corresponds to the K2~dwarf mid-eclipse (i.e., when the white dwarf is in front of the K2 star). Moreover, because the K2~dwarf rotates nearly synchronously, its rotational cycle is equal to the orbital cycle $E$.

\begin{figure}
    \centering
    \includegraphics[width=0.48\textwidth]{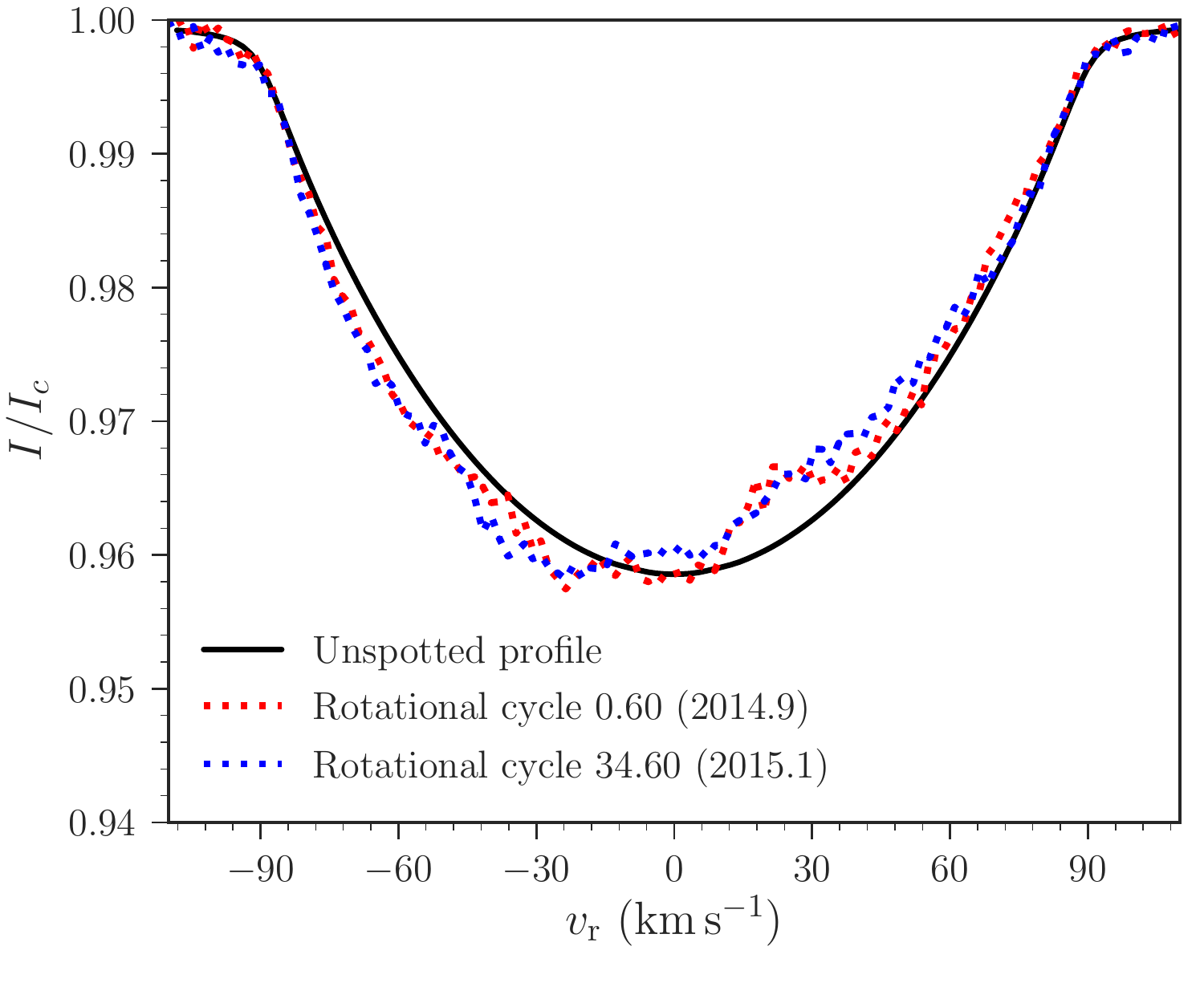}
    \caption{Observed Stokes $I$ LSD profiles at the rotational phase 0.6 (dotted lines) and absorption line profile computed for an unspotted star with $v\sin{i} = 89.3 \pm 0.1$\,km\,s$^{-1}$ (black continuous line). The two profiles correspond to observations at the same rotational phase in 2014.9 (red) and 2015.1 (blue). Rotational cycles (starting from cycle 21470) are indicated.}
    \label{fig:stokesIuns}
\end{figure}
\begin{figure*}
    \centering
    \includegraphics[width=2\columnwidth]{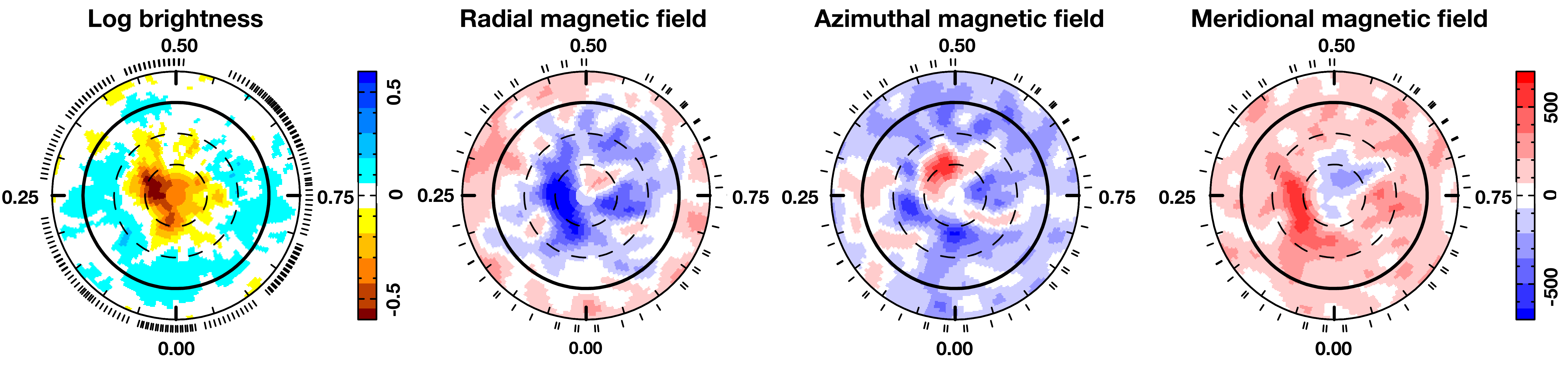}
    \caption{Tomographic reconstruction of the K2~dwarf surface using both 2014.9 and 2015.1 data sets. The star is shown in a flattened polar view with concentric circles representing $30\degr$ steps in latitude. Ticks outside the star indicate the rotational phase of our observations. The first plot shows the brightness distribution, where cool spots are shown as brown shades and warm plages as blue shades. The following plots show respectively the radial, azimuthal, and meridional components of the large-scale magnetic field in spherical coordinates. Magnetic fields are expressed in Gauss, with positive values represented in red shades and negative in blue.}
    \label{fig:map14and15}
\end{figure*}

In order to generate averaged photospheric lines of the K2~dwarf with enhanced SNRs, we applied Least-Squares Deconvolution \citep[LSD][]{DSC97} on all absorption lines with a relative depth of at least $40\%$ with respect to the continuum. Figure~\ref{fig:stokesIuns} shows two example Stokes $I$ LSD profiles obtained at rotational phase 0.6 using the same K2~dwarf absorption line mask detailed in \citetalias{ZDK21}. Observed Stokes $I$ LSD profiles show clear distortions with respect to the absorption line shape of an unspotted stellar surface assuming a line-of-sight projected equatorial velocity of $v\sin{i} = 89.3 \pm 0.11$\,km\,s$^{-1}$ (\citealt{VWV15}, \citetalias{ZDK21}). These Stokes $I$ signatures provide evidence for brightness inhomogeneities at the surface of the K2~dwarf star (i.e., signatures within $\pm v\sin{i}$), similar to what was found in previous Doppler images of this star (\citealt{RHJ95,HAS06}; \citetalias{ZDK21}; \citealt{KKO21}). Moreover, the shape difference of line profiles collected at the same rotational phase but different rotation cycles suggests that the brightness distribution evolves on a timescale of a few weeks.


\section{Results} \label{sec:results}
We apply the Zeeman-Doppler imaging technique (\verb'ZDI') to the time series of Stokes $I$ or Stokes $V$ LSD profiles to simultaneously reconstruct the surface brightness distribution and the large-scale magnetic field topology. To do so, \verb'ZDI' models the stellar surface as a grid of a few thousand cells, whose individual contributions to the total synthetic Stokes profiles are computed using the analytical solution of Unno-Rachkovsky to the polarised radiative transfer equations in a Milne-Eddington atmosphere \citep[see][]{LDL04}. The \verb'ZDI' code inverts the observed LSD profiles into surface images using a conjugate gradient algorithm that searches for the maximum-entropy image that reproduces the data down to a reduced $\chi^2$ of about unity \citep{DSP89,BDR91,DB97,DHJ06}. The entropy of each image is computed considering individual cells for the brightness maps, while it is a function of spherical harmonics coefficients for the magnetic maps. As in \citetalias{ZDK21}, the magnetic field expansion is limited to spherical harmonics with order $\ell \leq 15$. 

Our tomographic reconstruction follows closely the procedures described in \citetalias{ZDK21}, where we reconstructed brightness and large-scale magnetic surface maps of the K2~dwarf V471 Tau at two early epochs (2004.9 and 2005.9). In a first step, we use \verb'ZDI' to optimise the orbital motion correction by reconstructing surface spots from our set of Stokes $I$ profiles. Using a fixed semi-amplitude of $K = 149.3$\,km\,s$^{-1}$ \citepalias{ZDK21}, we reconstruct several brightness surface maps by varying the systemic velocity ($v_\gamma$) and phase offset ($\phi_0$) assuming the ephemeris of Equation~\ref{eq:ephemeris}. We find that, at constant information at the surface of the star, the best parameters reproducing the observations are $v_\gamma = 35.0 \pm 0.10$\,km\,s$^{-1}$ and $\phi_0 = 0.0025 \pm 0.0005$.

In all the image reconstructions that follow, we use the orbital parameters derived above to correct the spectra from Doppler shifts before applying \verb'ZDI'. Akin to \citetalias{ZDK21}, we fix the \BZ{line-of-sight} projected equatorial velocity and the stellar inclination angle to $v\sin{i} = 89.3 \pm 0.11$\,km\,s$^{-1}$ and $i = 78.755 \pm 0.030\,\degr$ \citep{VWV15}, respectively. We also recall that, due to the difficulty of \verb'ZDI' to distinguish features from Northern and Southern hemispheres in nearly equator-on stars \citep{VPH87,RWK89,UC95,SBM18,LHJ19,HKA21}, the reconstructed images may be subject to some mirroring effect with respect to the equator, to the same extent as those presented in \citetalias{ZDK21}. In practice, this effect is expected to be mitigated by the excellent phase coverage of our data set \citep[e.g. see][]{VPT93}.

\subsection{Brightness and magnetic imaging}\label{sec:maps}
We first attempt at reconstructing the surface maps of the K2~dwarf star using the LSD profiles collected in 2014.9 and 2015.1. Applying \verb'ZDI' to the Stokes $I$ LSD profiles (Stokes $V$ LSD) shows that the data can only be fitted down to a reduced $\chi^2$ of 1.47 (1.15) when assuming that the star rotates as a solid body. When assuming differential rotation (see Section~\ref{sec:dr}), the Stokes $I$ data can now be fitted down to a reduced $\chi^2$ of $1.10$ and Stokes $V$ data to $1.07$.

Figure~\ref{fig:map14and15} shows the maps obtained after including differential rotation in our image reconstruction process. The brightness map recovered for the combined 2014.9 and 2015.1 data set shows inhomogenieties with respect to the unperturbed photosphere (with an effective temperature of about 5066\,K). It features a cool polar cap with low-latitude appendages that extend down to $30\degr$ latitude. As in \citetalias{ZDK21}, we find that warm low-contrast plages forming a partial ring structure are also present at low latitudes. From the brightness map we obtain that $10\%$ of the stellar surface is covered with cool spots and $8\%$ with warm plages.

The reconstructed large-scale magnetic field is also shown in Figure~\ref{fig:map14and15}. We find an average magnetic field strength of 360\,G. It shows up from the surface maps that strong negative radial fields (reaching strengths up to 500\,G) overlap with the high-contrast cool spots forming the polar cap. The overall magnetic topology that we obtain is dominated by the poloidal component, whereas the toroidal magnetic energy accounts for $25\%$ of the total energy. The poloidal field features a strong dipole mode (containing $60\%$ of the poloidal energy), while other spherical harmonics modes with order $\ell \geq 4$ contribute altogether to $30\%$ of the poloidal energy. We also find that $75\%$ of the poloidal energy is stored in axisymmetric modes with $m<\ell/2$. The dipolar component has a polar strength of 335\,G and is tilted by $7\degr$ towards phase 0.87. 

\pagebreak

\subsubsection{Short-term variability} \label{sec:var}
We find that even after including differential rotation in our image reconstruction process, the total data set can only be fitted down to a reduced $\chi^2$ of 1.1 when using Stokes $I$ profiles and 1.07 when using Stokes $V$. This suggests that the surface brightness and magnetic maps evolve on a timescale of a few weeks. To explore whether a short-term evolution indeed occurs, we split the total data set in two. The first subset gathers spectra from 2014.9 (totalling 132 Stokes $I$ profiles and 33 Stokes $V$ spread over 6 non-consecutive nights) and the other combines spectra from 2015.1 (104 Stokes $I$ profiles and 26 Stokes $V$ collected over 5 nights).

\begin{figure}
    \centering
    \includegraphics[width=\columnwidth]{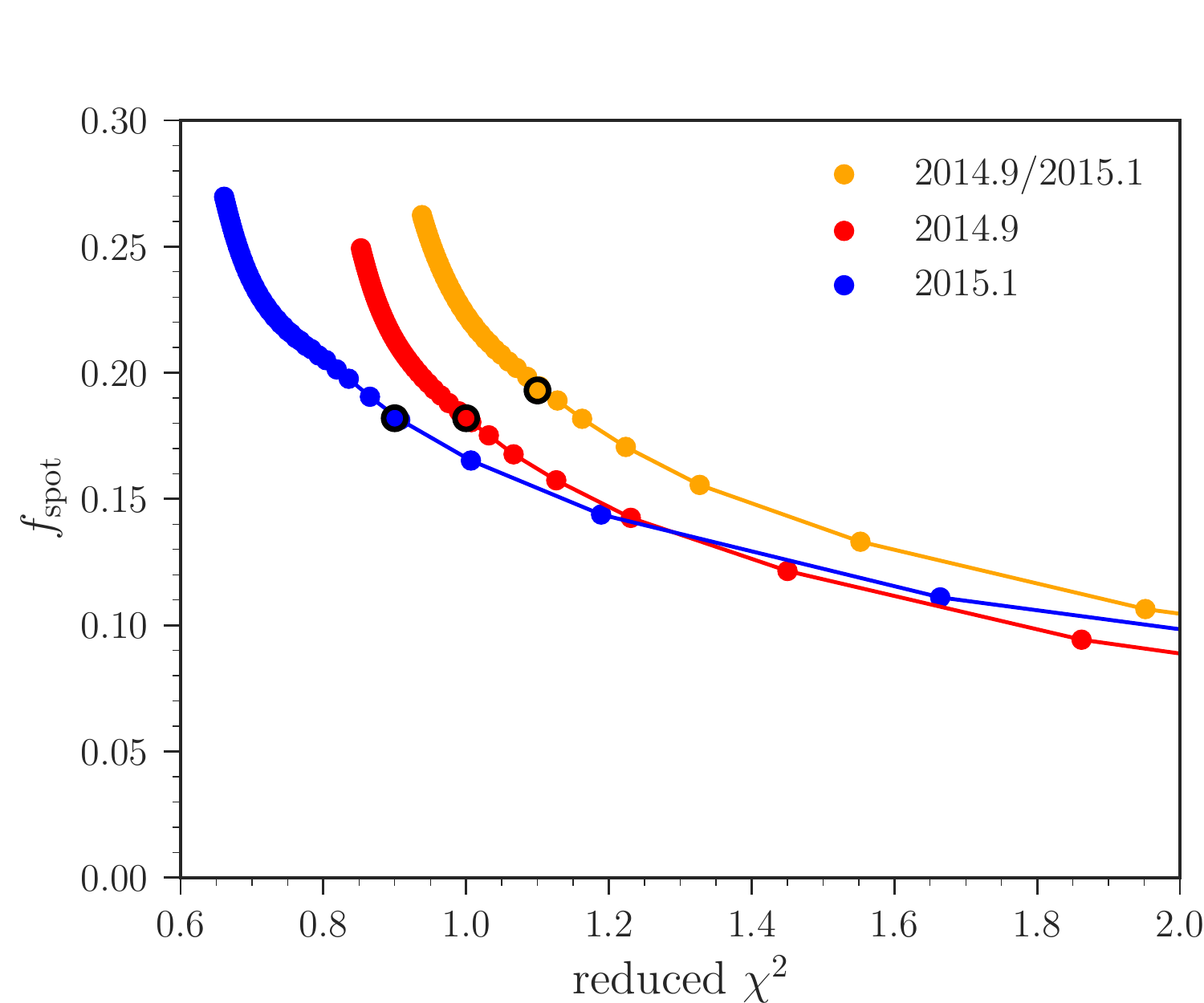}
    \caption{Spot coverage ($f_\text{spot}$) as a function of the reduced $\chi^2$ for a tomographic reconstruction process aiming at a low value of reduced $\chi^2$ and including differential rotation (see Section~\ref{sec:dr}). Curves with different colors show reconstructions using different Stokes $I$ data sets (see legend). Black circles highlight the values when the spot coverage sharply rises for a decreasing reduced $\chi^2$.}
    \label{fig:caim}
\end{figure}

Figure~\ref{fig:caim} shows the spot coverage at different iterations of the  \verb|ZDI| reconstruction process when we aim at fitting the time series of Stokes $I$ profiles at a progressively lower reduced $\chi^2$. The data illustrates how \verb|ZDI| adds spots at the stellar surface to better fit the observations. We can see in this figure that the spot coverage sharply increases below a given reduced $\chi^2$ (whose value depends on the data set). This behavior suggests that below this reduced $\chi^2$ threshold, the tomographic imaging process starts to fit noise features present in the data. Using the slope of the curves in Figure~\ref{fig:caim} as a criteria to define the reduced $\chi^2$ at which the \verb|ZDI| reconstruction process aims \citep[e.g., see][for a detailed explanation about using derivatives as a stop-criteria]{AHG15}, we find that the Stokes $I$ subsets of 2014.9 and 2015.1 can be fitted down to a reduced $\chi^2$ of 1.0 and 0.9, respectively.

\begin{figure*}
    \centering
\includegraphics[width=2\columnwidth]{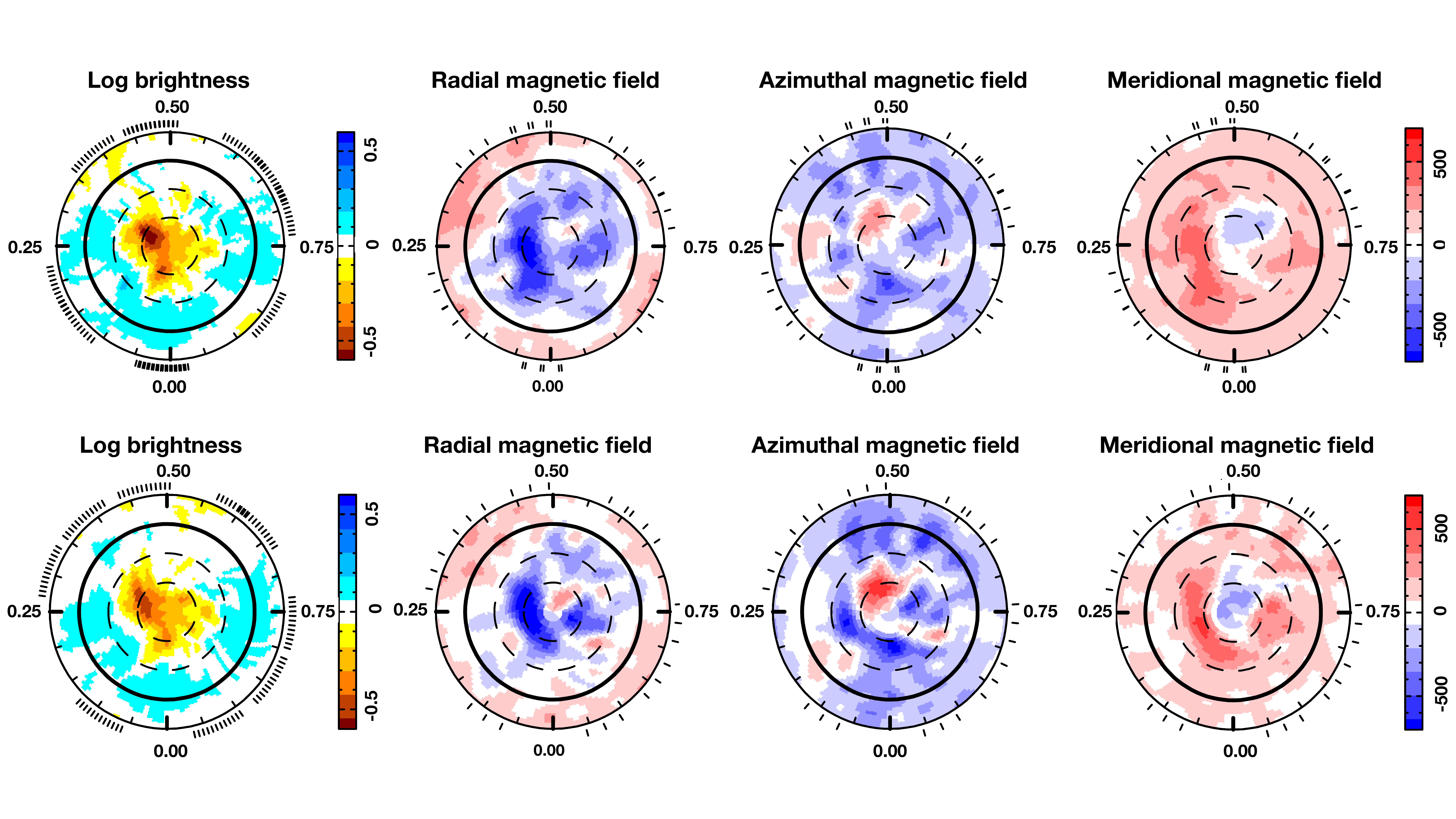}    
    \caption{ZDI reconstruction of the individual subsets of 2014.9 (top panels) and 2015.1 (bottom panels). Surface maps are illustrated in a similar fashion to Figure~\ref{fig:map14and15}.}
    \label{fig:mapsplit}
\end{figure*}

Figure~\ref{fig:mapsplit} shows the brightness and magnetic maps obtained for the individual subsets including differential rotation.  Whereas the maps derived in 2014.9 and 2015.1 look similar at first order, we observe small differences reflecting an intrinsic evolution of the brightness and magnetic field of the K2~dwarf. Starting from the brightness maps, we note a visible decrease in the contrast of the cool spot at the polar cap and tiny azimuthal rearrangements in the distribution of warm plages. As a result, the K2~dwarf surface appears slightly less spotted in 2015.1. We find that warm plages cover $7\%$ of the surface in both maps, while dark spots covered $10\%$ of the stellar surface in 2014.9 and $9\%$ in 2015.1. Regarding magnetic maps, we find that the negative radial field covers a larger portion of the North pole in 2014.9 than 2015.1 (see Figure~\ref{fig:mapsplit}).  As expected, the brightness and magnetic maps derived from the original data set (Figure~\ref{fig:map14and15}) resemble an average of the individual maps derived from the subsets. Table~\ref{mag_tab} summarises the magnetic properties derived from the split and original data sets. We assess the uncertainties in the image reconstruction process using the bootstrap technique detailed in \citetalias{ZDK21}.

\begin{table}
	\centering
	\caption{Magnetic field properties of the K2~dwarf star. B$_\mathrm{rms}$ is the root-mean-square field, B$_\mathrm{dip}$ is the dipolar strength, and E$_\mathrm{pol}$ is the fractional energy in the poloidal field. E$_{\ell = 1}$, E$_{\ell = 2}$, E$_{\ell = 3}$ and E$_{\ell \geq 4}$ are, respectively, the fractional energies of the dipolar, quadrupolar, octupolar, and multipolar (defined as $\ell \geq 4$) components.}
	\label{mag_tab}
	\begin{tabular}{cccc} 
	 \cline{1-4}
	 & \multicolumn{3}{c}{Data set} \\
	 & 2014.9/2015.1 & 2014.9 & 2015.1 \\
	 \hline
	B$_\mathrm{rms}$ (G) & $ 360 \pm 7$  & $ 415 \pm 5$ & $ 335 \pm 6$ \\
	B$_\mathrm{dip}$ (G) & $- 335 \pm 20$ & $- 440 \pm 35$ & $- 250\pm 26$\\
	$\theta_\mathrm{dip} \left(\degr \right)$ & $ 7 \pm 5$  & $9 \pm 2$ & $9 \pm 3$\\
	E$_\mathrm{pol} (\%)$ & $ 75 \pm 5$ & $ 80 \pm 5$ & $ 70 \pm 3$\\
	E$_{\ell = 1} (\%)$  & $  55 \pm 5$ & $ 60 \pm 9$ &  $ 45 \pm 7$\\
	E$_{\ell = 2} (\%)$  & $ 5 \pm 5$ & $5 \pm 2$ & $5 \pm 2$ \\
	E$_{\ell = 3} (\%)$  & $ 5 \pm 2$ & $5 \pm 2$ & $5 \pm 2$ \\
	E$_{\ell \geq 4 } (\%)$ & $ 35 \pm 4$ & $ 30 \pm 9$ & $ 45 \pm 7$\\
	\hline
	\end{tabular}
\end{table}

\subsection{Differential rotation} \label{sec:dr}
As mentioned in Subsection~\ref{sec:maps}, the global data set 2014.9/2015.1 cannot be fitted down to a reduced $\chi^2$ of 1 as a result of temporal evolution of surface maps. One of the potential sources for this evolution is the presence of differential rotation at the surface of the K2~dwarf star.

The \verb'ZDI' code allows one to explore whether stars rotate differentially by searching for recurrent distortions in the line profiles of our spectropolarimetric time series \citep{DMC00}. To do so, \verb'ZDI' incorporates in the image reconstruction process a predefined latitudinal differential rotation law given by
\begin{equation}
    \Omega(\theta) = \Oeq -\dO \sin^2(\theta),
\end{equation}
where $\Omega$ is the latitudinal angular velocity profile, $\theta$ is the latitude, $\Oeq$ is the angular velocity at the equator, and $\dO$ is the difference between $\Oeq$ and the angular velocity at the pole. Brightness and magnetic maps are thus individually reconstructed for each pair of ($\Omega_\mathrm{eq}, \mathrm{d}\Omega$) values, with a $\chi^2$ value attributed to each tomographic reconstruction (carried out at constant information content for all pairs of differential rotation parameters). 
\begin{figure}
    \centering
    \includegraphics[width=\columnwidth]{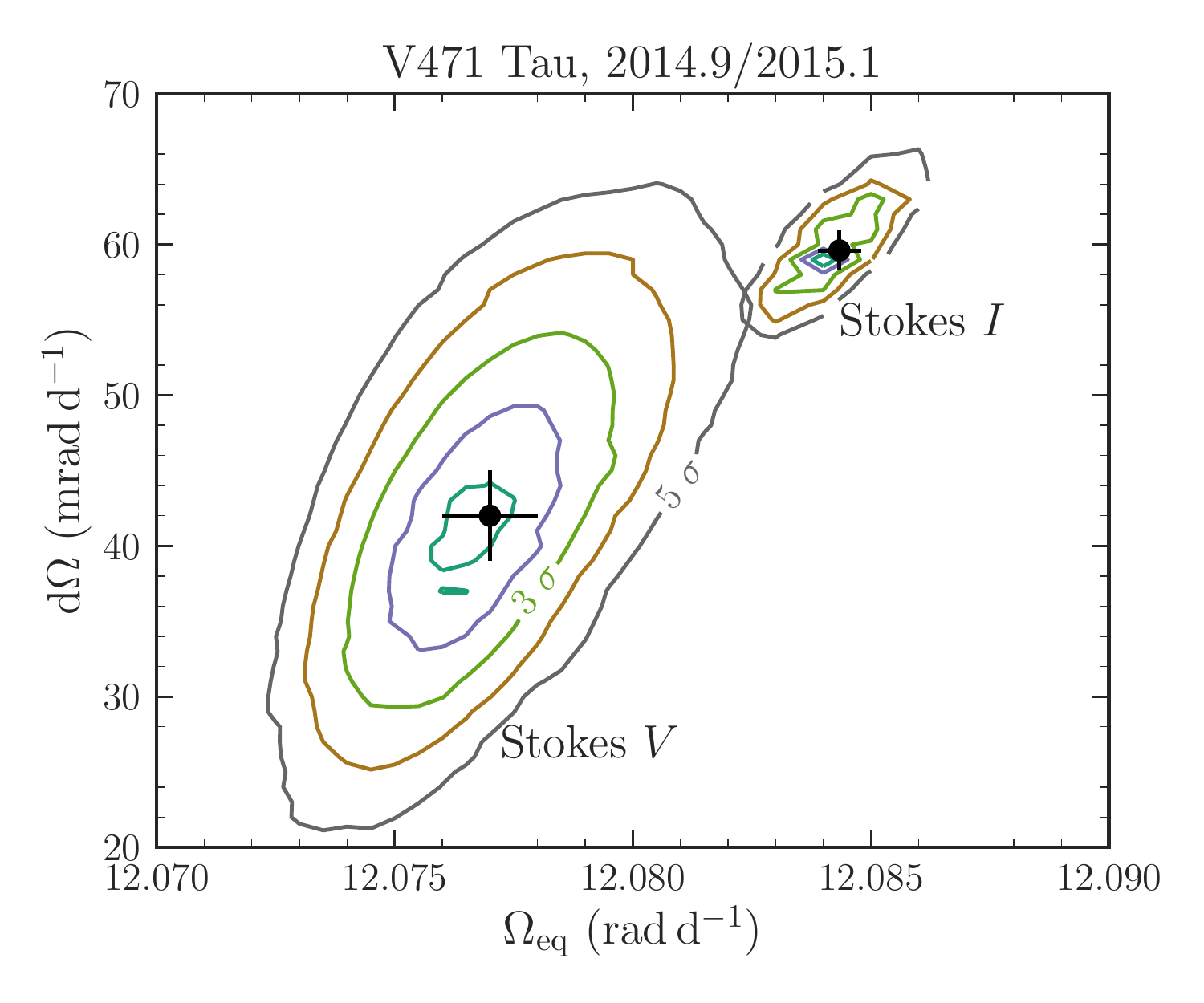}
    \caption{Differential rotation measurements obtained with the sheared-imaging method using Stokes $I$ or Stokes $V$ profiles. Contour levels represent confidence levels up to $5\,\sigma$. Black circles represent the best-fit obtained after assuming a paraboloid distribution.}
    \label{fig:DR}
\end{figure}
Figure~\ref{fig:DR} shows the resulting confidence levels for the differential rotation parameters, when reconstructing the brightness surface distribution (i.e., using Stokes $I$ alone) and the magnetic topology (i.e., using Stokes $V$ alone). Using the minimum of the paraboloid and its local curvature to retrieve the optimal shear parameters and corresponding error bars \citep{DCP03}, we obtain $\dO = 60 \pm 1 $\,mrad\,d$^{-1}$ and $\Oeq = 12.084 \pm 0.001$\,rad\,d$^{-1}$ from the $\chi^2$ distribution using Stokes $I$, and $\dO = 42 \pm 3$\,mrad\,d$^{-1}$ and $\Oeq = 12.077 \pm 0.001 $\,rad\,d$^{-1}$ from the $\chi^2$ distribution using Stokes $V$. As in \citetalias{ZDK21}, these results again suggest that brightness inhomogeneities and magnetic structures are sheared by different amounts. 

Moreover, we use ten bootstrapped data sets to estimate the differential rotation parameters uncertainty associated with the image reconstruction process (see details in \citetalias{ZDK21}). To do that, we repeat the steps above to determine the differential rotation parameters using bootstrapped Stokes $I$ or Stokes $V$ profiles. From the $\chi^2$ maps obtained using bootstrapped data sets, we find that the mean values of the ten error bars obtained using Stokes $I$ and Stokes $V$ profiles are similar to the error bars obtained from the original data.

\subsection{H\texorpdfstring{$\alpha$}{alpha} variability}

\begin{figure}
    \centering
    \includegraphics[width=\columnwidth]{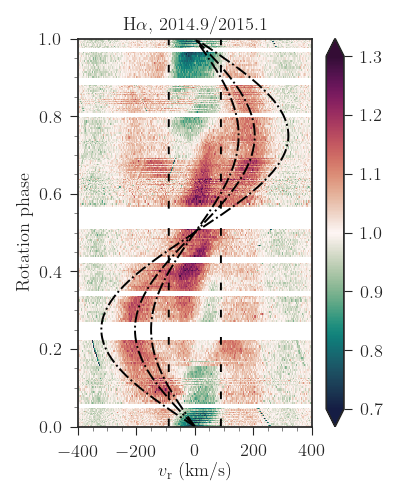}
    \caption{Dynamical spectra of H$\alpha$ shown in the rest frame of the K2 dwarf. Red shades indicate that H$\alpha$ is in emission and green shades in absorption. The vertical dashed lines correspond to the stellar rotational broadening of $\pm v\sin(i)$. Sine waves of \BZ{semi-amplitudes} $150$\,km\,s$^{-1}$ (center of mass), $205$\,km\,s$^{-1}$ (prominence position), and $320$\,km\,s$^{-1}$ (white dwarf position) are over-plotted as dashed-dotted lines. }
    \label{fig:halpha}
\end{figure}
The variability of H$\alpha$ in V471~Tau has been reported by several authors (\citealt{YRS91,RBY02,KKO21}, \citetalias{ZDK21}). The dynamical spectra of the H$\alpha$ line in 2014.9/2015.1 is plotted in Figure~\ref{fig:halpha}. Starting with the radial velocity range within $\pm v\sin(i)$, we identify that H$\alpha$ exhibits the typical rotational modulation at the the K2 dwarf surface. H$\alpha$ is in emission on the stellar hemisphere that faces the white dwarf and in absorption on the opposite hemisphere. The H$\alpha$ equivalent width reveals a peak-to-peak amplitude of about 1.2\,{\AA} with a maximum emission of $-0.5$\,{\AA} at phase 0.5.

Further, we observe a modulated emission with an amplitude of $205 \pm 40$\,km\,s$^{-1}$ in the rest frame of the K2~dwarf (see Figure~\ref{fig:halpha}). We speculate that this emission is due to a stable prominence trapped at $2.30 \pm 0.45$\,R$_\star$ from the K2~dwarf (or, equivalently, at $1.29 \pm 0.45$\,R$_\star$ from the white dwarf component). We find a full width at half maximum (FWHM) of $1.89$\,{\AA} and an equivalent width of about $-0.17$\,{\AA} when fitting a Gaussian to the prominence emission at phase 0.75. Assuming that the prominence is spherical, we estimate a prominence radius of about $0.50$\,R$_\star$ from its FWHM emission in H$\alpha$.


\section{Discussions and conclusions} \label{sec:conclusions}

In this paper, we analysed new spectropolarimetric data of the binary system V471~Tau collected from 20 December 2014 to 12 January 2015 with \verb|ESPaDOnS|. Using Zeeman-Doppler imaging, we modelled time series of LSD Stokes $I$ and $V$ profiles to recover new brightness and magnetic maps of the K2~dwarf component of V471 Tau.

\subsection{Brightness map, magnetic field topology, and differential rotation}
Our brightness image reveals a strong cool polar cap in 2014.9/2015.1. This result is confirmed by an independent Doppler imaging reconstruction using a different inversion code \citep[][]{KKO21}. Along with previous brightness maps (\citealt{RHJ95,HAS06}; \citetalias{ZDK21}), we find that the cool polar cap seen in the star surface is stable in a timescale of years. The spot coverage of $\approx 18\%$ derived in 2014.9/2015.1 is in good agreement with what is expected from photometry \citepalias[in the range 15–25$\%$, see][]{ZDK21} suggesting that most of the brightness spots generating photometric fluctuations in V471~Tau are large enough to be detected and resolved by Doppler imaging.

The reconstructed large scale magnetic field shows a dominant poloidal component that accounts for about $75\%$ of the magnetic energy in 2014.9/2015.1. This value is slightly larger than those observed in 2004.9 and 2005.9, whose fractional poloidal energy corresponded to $70\%$ and $60\%$, respectively. Moreover, we find that the dipole strength in 2014.9/2015.1 is about 3.6 times stronger than that in 2004.9 and 2005.9. 

We also confirmed that the surface of the K2~dwarf is differentially rotating. We measured an equatorial to pole angular velocity difference of 60 and 42\,mrad\,d$^{-1}$ from spot and magnetic structures, respectively. This finding confirms the solar-like differential rotation profile obtained for the star in 2004.9 and 2005.9 \citepalias{ZDK21}. Interestingly, the shear level inferred from our 2014.9/2015.1 data set resembles closely those obtained nine years before \citepalias[73 and 48 \,mrad\,d$^{-1}$ in 2005.9;][]{ZDK21}.

\subsection{Magnetic activity}
Studies of chromospheric/coronal activity indicators \citep{RBY02,KRJ07,PS08,KKO21} and long-term photometry \citep{SP88,IET05} of the K2~dwarf suggest an activity cycle of about 13\,yr. This possible activity cycle indicates that the two data sets analysed in \citetalias{ZDK21} (2004.9 and 2005.9) occurred at activity minimum (spanning from late-2004 to late-2007), whereas the data set analysed in this paper (2014.9/2015.1) took place close to activity maximum (spanning from late-2011 to late-2014). Such scenario is indeed corroborated by our large-scale magnetic field maps. We find that the averaged unsigned magnetic field strength increased by about 2.2 times from the two first epochs (at activity minimum) to the last epoch (at activity maximum). No such modulation is visible in the brightness maps, which display a spot coverage of $14\%$, $17\%$, and $18\%$ in 2004.9, 2005.9, and 2014.9/2015.1, respectively. This result emphasizes that spot coverage may not always be an appropriate observable to study activity cycles in very active rapidly rotating stars.

The analysis of the H$\alpha$ emission in 2014.9/2015.1 shows a prominence located farther than the Lagrange point L1 towards the white dwarf component, and that remained stable during the entire observation window (44 rotation cycles). The prominence size and location we infer are consistent within error bars with the prominence properties derived in 2004.9 \citepalias{ZDK21}. Using the prominence flux in H$\alpha$ of $1.1 \times 10^{-13}$\,erg\,s$^{-1}$\,cm$^{-2}$, we derive a prominence mass of $4\times10^{18}$\,g in 2014.9/2015.1 slightly smaller than that identified in 2004.9 of $6\times10^{18}$\,g \citep[see Equation~3 in][]{SHM96}. The prominence mass-range of 4–6$\times10^{18}$\,g is broadly consistent with those derived for other K~dwarf stars hosting prominences – e.g. K0 dwarf AB~Dor \citep[2-10$\times 10^{17}$\,g, e.g., ][]{CR89,CDE90} and the K3 dwarf Speedy Mic \citep[0.5–2.3$\times10^{17}$\,g,][]{DCB06}.

\begin{figure}
	\centering
	\includegraphics[width=\columnwidth]{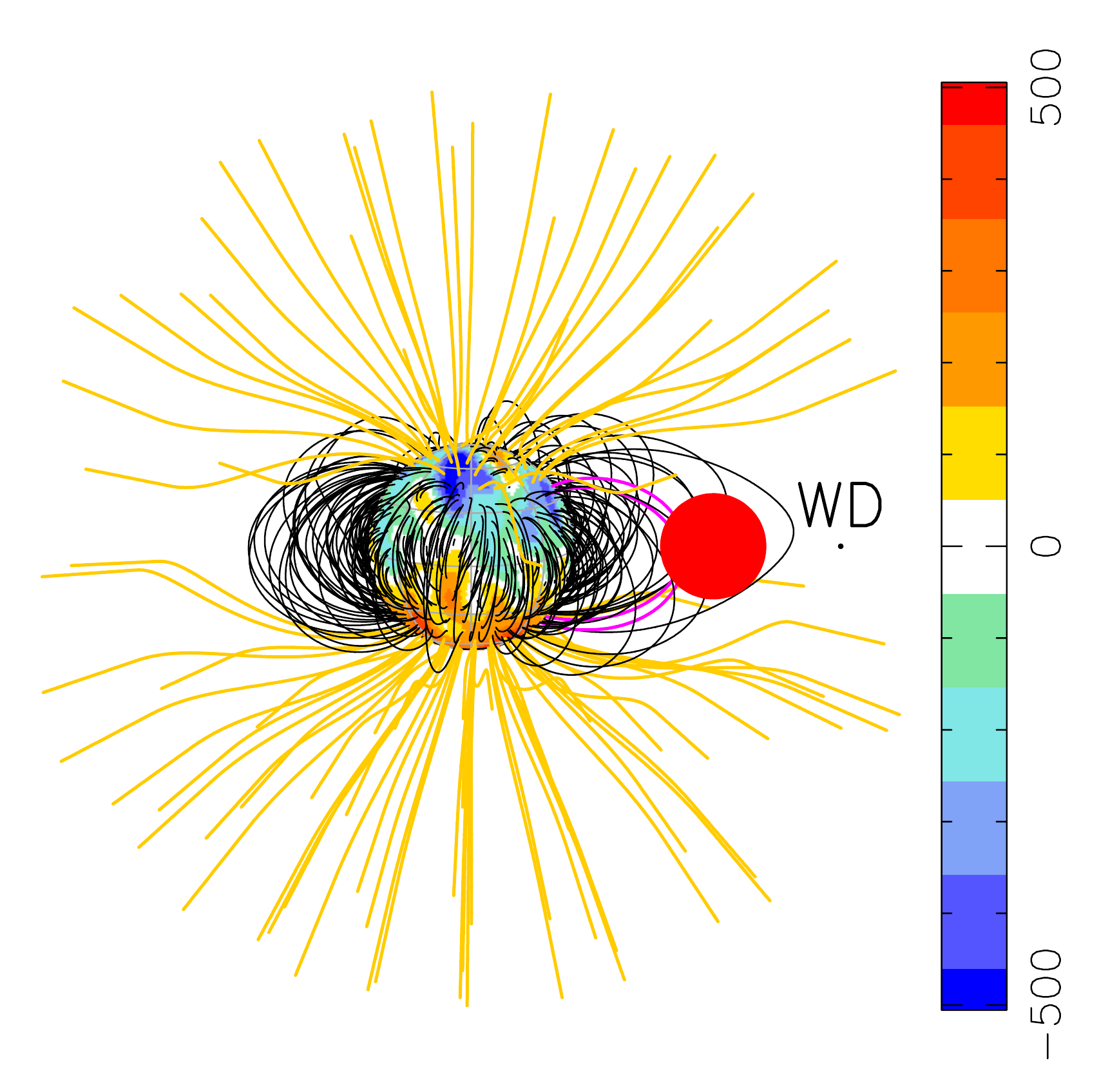}
    \caption{Potential field extrapolation of the large-scale radial magnetic field reconstruction of the K2~dwarf obtained with ZDI in 2014.9/2015.1. Field lines are seen at rotational phase 0.75 and are shown in yellow/black when the lines are open/closed. The prominence is illustrated as a red circle, and the field lines crossing the prominence are coloured in magenta. The local surface field strength (G) of the star is shown in colours and follows the colour scale on the right. A black circle indicates the white dwarf (WD) position; however, its magnetic field is not considered in the potential field extrapolation. 
    }
    \label{fig:potfield}
\end{figure}
As illustrated in Figure~\ref{fig:potfield}, the potential field extrapolation of the radial magnetic field map that we derived for the K2~dwarf shows closed loops of magnetic field that extend out from the surface and reach the prominence location. This result is consistent with that of \citetalias{ZDK21} and offers further qualitative proof that a slingshot mechanism is likely responsible for confining the prominence further away from the center of mass of the system and from the Lagrange point L1 (located at $1.679\pm0.004\,\mathrm{R_\star}$ and $1.84\pm0.02\,\mathrm{R_\star}$ from the center of the K2~dwarf star, respectively). Slingshot mechanisms have been also suggested to operate in single fast-rotating stars hosting prominences at a few stellar radii above the stellar surface \citep[see discussion in][]{JCC19}, such as AB~Dor \citep{CR89,WJ19}, HK Aqu \citep{BER96}, LQ~Lup \citep{DMC00}, Speedy Mic \citep{DCB06,WJ19}, V374 Peg \citep{VKO16}, and V530 Per \citep{CPD20,CPD21}.

\begin{figure*}
	\centering
	\includegraphics[width=\textwidth]{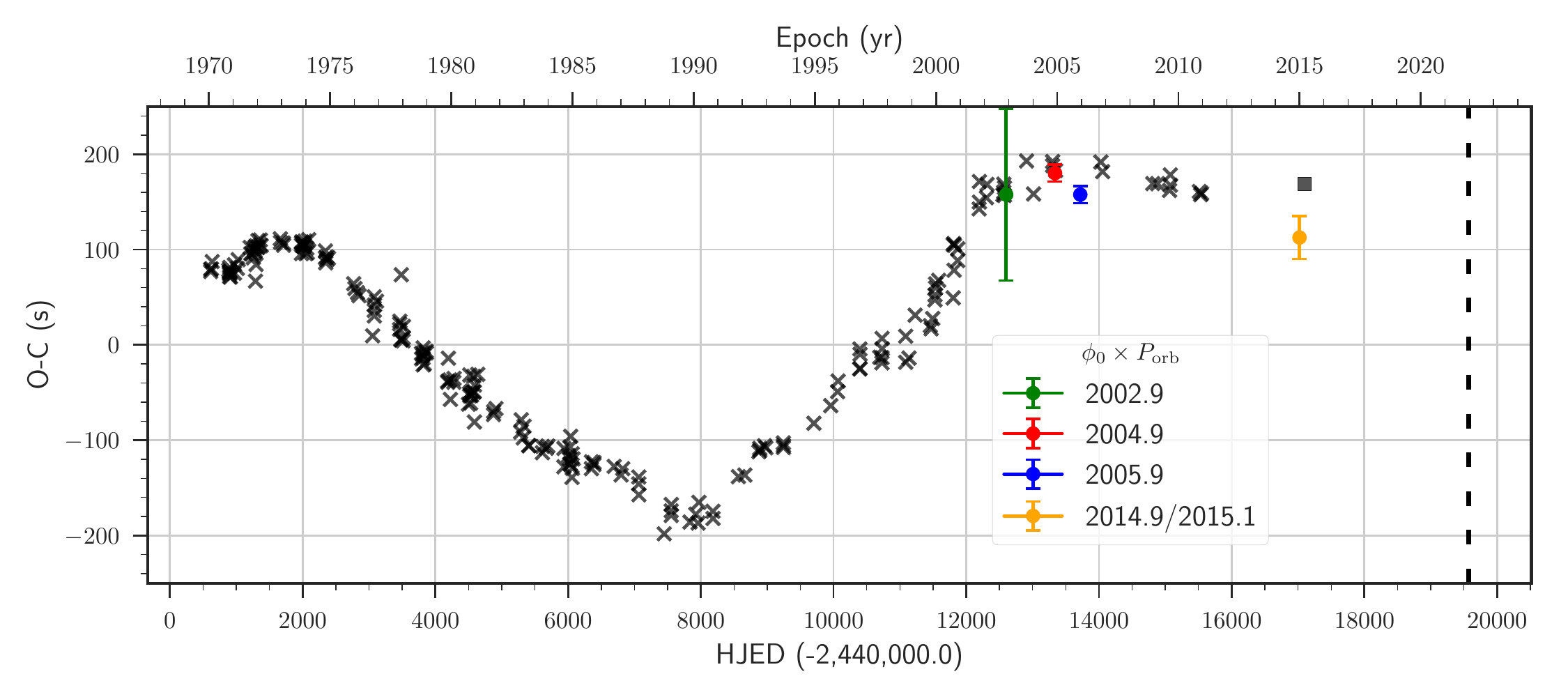}
    \caption[Observed minus computed eclipse timings]{Observed minus computed eclipse timing variations of V471 Tau assuming the linear ephemeris given by Equation~\ref{eq:ephemeris}.
    Crosses represent O-C measurements from eclipse timings and circles the O-C values estimated from the phase offset $\phi_0$ multiplied by the orbital period (shown with 1-$\sigma$ errorbars).
    The black square gives the single eclipse timing measure using K2 data \citep[][]{MND21}, corresponding to HJD 2457097.1816484 in Terrestrial Time (TT) scale \citep[see][]{ESG10}. The vertical dashed line marks the \BZ{recent observation campaign of V471~Tau with ESPaDOnS (2021B)}.}
    \label{fig:ocplot}
\end{figure*}

Altogether, the potential field extrapolations available for the K2~dwarf V471~Tau show that when a prominence is seen in the system (2004.9 and 2014.9/2015.1) close loops of magnetic lines reach the prominence location, whereas when no prominence is detected (2005.9) only open field lines are found at the expected prominence location (see details about previous reconstructions in \citetalias{ZDK21}). This finding indicates that the evolution of the large-scale magnetic field controls the rate at which stable prominences are generated in V471~Tau.

\subsection{ETVs in V471~Tau}

The observed minus computed (O-C) eclipse timings available in the literature for V471 Tau (cross symbols) are illustrated in Figure~\ref{fig:ocplot}. The data evidence the periodic behaviour of the ETVs in V471 Tau with current observations yielding a modulation period of 30-40\,yr \citep{GR01,IET05,KH11,MGE18} and an O-C amplitude ranging from 130 to 200\,s \citep{KH11,MGE18} depending on the ephemeris employed. To compare this trend with independent measures, we use the phase offsets $\phi_0$ available in the literature for V471~Tau to infer the O-C amplitude. \BZ{These} two quantities are expected to scale as
\begin{equation} \label{oc_phase}
    \text{O-C} = \phi_0\times P_\mathrm{orb}.
\end{equation}
We find an O-C amplitude of $158 \pm 90$\,s in 2002.9 \citep[green circle,][]{HAS06}, $180 \pm 9$\,s in 2004.9 \citepalias[red,][]{ZDK21}, $158 \pm 9$\,s in 2005.9 \citepalias[blue,][]{ZDK21}, and $113 \pm 23$\,s in 2014.9/2015.1 (yellow, this paper). 

The O-C values that we infer from Equation~\ref{oc_phase} agree with the trend found using long-term photometry. They offer an independent validation of the ETVs in V471~Tau as all the four phase offset measurements considered take into account the presence of spots at the surface of the K2~dwarf (see Section~\ref{sec:results}) that may otherwise affect the eclipse timing measurements from photometry \citep{KRL02}. Furthermore, the intermediate value of O-C that we infer in 2014.9/2015.1 suggests that although the orbital period of the system was decreasing from 2002.9 to 2015.1, it did not reach the minimum orbital period recorded for V471 Tau (which took place around 1980, i.e. when O-C crosses 0 going to negative values). As a result, the observation reported in this paper did not probe the ETV cycle at the phase of largest surface differential rotation as predicted if an Applegate mechanism is indeed operating on V471 Tau. This result is corroborated by the eclipse timing measure using photometric data from the $K2$ mission acquired around 2015.2, which yields O-C = 169\,s.

Similar to the findings of \citetalias{ZDK21}, we detect a relative differential rotation in 2014.9/2015.1 of $\dO/\Oeq = 0.5\%$ and $0.4\%$ using Stokes $I$ and Stokes $V$, respectively. These values are weaker than what is needed for the feasibility of an Applegate mechanism in V471~Tau. Considering the range of shears currently measured at the surface of the K2~dwarf (ranging from 0.4 to 1.1$\%$), the Applegate mechanism would drive ETVs with a semi-amplitude of $\Delta P/P_\mathrm{orb} \lesssim 10^{-7}$ \citep{VSB18} whereas V471~Tau displays $\Delta P/P_\mathrm{orb}~\approx~8.5\times10^{-7}$. Nevertheless, our O-C estimation in 2014.9/2015.1 indicates that the system was not orbiting at the minimum orbital period expected for V471~Tau (i.e., when the largest surface shear is expected in the framework of the Applegate mechanism). It may be possible that higher values of $\dO/\Oeq$ occur at the surface of the K2~dwarf and thus that the Applegate mechanism may indeed be at work.

We suggest that spectropolarimetric observations in the upcoming years will help understand whether the ETVs in V471~Tau are magnetically-driven especially if they can probe the ETV cycle at the expected phase of largest differential rotation. Along with the tomographic maps already reconstructed for the K2~dwarf, it will be possible to further investigate whether the ETVs of V471 Tau are caused by the mechanism proposed by \citet{A92} or by \citet{L20}. \BZ{For this purpose, observations of V471~Tau were recently collected with ESPaDOnS in 2021B}.


\section*{Acknowledgements}
We thank the anonymous referee for helping improve and clarify this manuscript. This project received funding from the European Research Council (ERC) under the H2020 research $\&$ innovation programme (grant agreements $\#740651$ New-Worlds and $\#865624$ GPRV). This paper is based on observations obtained at the Canada-France-Hawaii Telescope (CFHT) which is operated by the National Research Council of Canada, the Institut National des Sciences de l'Univers of the Centre National de la Recherche Scientique of France, and the University of Hawaii. The observations at the CFHT were performed with care and respect from the summit of Maunakea which is a significant cultural and historic site.

\section*{Data Availability}
This paper includes data collected by the ESPaDOnS spectropolarimeter, which is publicly available from the Canadian Astronomy Data Center (program IDs: 15AP15 $\&$ 14BP15).



\bibliographystyle{mnras}
\bibliography{ref} 



\appendix
\section{Journal of observations}\label{sec:tables}

The logbook of the spectropolarimetric observations of V471~Tau used in this study is shown in Table~\ref{tab:Data}. 

\begin{table*}	
\setlength{\tabcolsep}{4.4pt}
	\caption{Summary of ESPaDOnS/CFHT observations for V471~Tau from December 2014 to January 2015. Columns 1 to 4 respectively record  (i) the date of observation, (ii) the UT time at mid sub-exposure, (iii) the time in Heliocentric Julian Date (HJD), and (iv) the rotation cycle of each observation. Column 5 illustrates peak SNR values for the Stokes $V$ spectrum (per $1.8$~km/s spectral pixel). Column 6 shows the RMS noise level of the Stokes $V$ LSD profile.}
	\begin{tabular}{lccccc} 
		\hline
		Date & UT        & HJD    & $E$   &   SNR &  $\sigma_{LSD}$  \\
		 & (h:m:s) & $(2,453,337+) $ & $(21,470+)$   &  & $(10^{-4}) $  \\
		\hline
20 Dec 2014  &  06:17:40 &  3674.76735  & 0.118441 & 202  & 1.5  \\
20 Dec 2014  &  06:35:24 &  3674.77967  & 0.142079 & 207  & 1.5  \\
20 Dec 2014  &  06:53:09 &  3674.79199  & 0.165718 & 200  & 1.5  \\
20 Dec 2014  &  12:05:19 &  3675.00877  & 0.581656 & 137  & 2.5  \\
20 Dec 2014  &  12:23:04 &  3675.02109  & 0.605294 & 134  & 2.5  \\
20 Dec 2014  &  12:40:48 &  3675.03340  & 0.628914 & 122  & 2.9  \\
21 Dec 2014  &  05:41:49 &  3675.74239  & 1.989260 & 194  & 1.6  \\
21 Dec 2014  &  05:59:33 &  3675.75471  & 2.012899 & 195  & 1.6  \\
21 Dec 2014  &  06:17:18 &  3675.76704  & 2.036556 & 193  & 1.6  \\
21 Dec 2014  &  11:29:48 &  3675.98404  & 2.452917 & 182  & 1.8  \\
21 Dec 2014  &  11:47:33 &  3675.99636  & 2.476555 & 179  & 1.8  \\
21 Dec 2014  &  12:05:17 &  3676.00868  & 2.500194 & 165  & 2.0  \\
22 Dec 2014  &  06:40:09 &  3676.78284  & 3.985582 & 187  & 1.7  \\
22 Dec 2014  &  06:57:53 &  3676.79515  & 4.009202 & 196  & 1.6  \\
22 Dec 2014  &  07:15:38 &  3676.80748  & 4.032859 & 198  & 1.6  \\
22 Dec 2014  &  12:26:57 &  3677.02366  & 4.447646 & 176  & 1.9  \\
22 Dec 2014  &  12:44:42 &  3677.03598  & 4.471285 & 178  & 1.9  \\
22 Dec 2014  &  13:02:26 &  3677.04830  & 4.494923 & 157  & 2.1  \\
28 Dec 2014  &  08:56:52 &  3682.87737  & 15.679221 & 205  & 1.5  \\
28 Dec 2014  &  09:14:36 &  3682.88968  & 15.702840 & 209  & 1.5  \\
28 Dec 2014  &  09:32:20 &  3682.90200  & 15.726479 & 184  & 1.7  \\
29 Dec 2014  &  06:02:07 &  3683.75595  & 17.364961 & 196  & 1.6  \\
29 Dec 2014  &  06:19:52 &  3683.76827  & 17.388600 & 198  & 1.6  \\
29 Dec 2014  &  06:37:36 &  3683.78059  & 17.412238 & 189  & 1.6  \\
29 Dec 2014  &  11:46:02 &  3683.99477  & 17.823188 & 194  & 1.6  \\
29 Dec 2014  &  12:03:47 &  3684.00709  & 17.846826 & 191  & 1.7  \\
29 Dec 2014  &  12:21:32 &  3684.01941  & 17.870465 & 182  & 1.8  \\
30 Dec 2014  &  04:34:15 &  3684.69486  & 19.166458 & 193  & 1.6  \\
30 Dec 2014  &  04:52:01 &  3684.70719  & 19.190115 & 196  & 1.6  \\
30 Dec 2014  &  05:09:45 &  3684.71951  & 19.213754 & 203  & 1.6  \\
30 Dec 2014  &  10:24:43 &  3684.93823  & 19.633414 & 210  & 1.5  \\
30 Dec 2014  &  10:42:28 &  3684.95055  & 19.657053 & 210  & 1.5  \\
30 Dec 2014  &  11:00:12 &  3684.96287  & 19.680691 & 212  & 1.5  \\
07 Jan 2015  &  05:43:50 &  3692.74255  & 34.607644 & 133  & 2.5  \\
07 Jan 2015  &  06:01:37 &  3692.75490  & 34.631341 & 138  & 2.4  \\
07 Jan 2015  &  06:19:24 &  3692.76725  & 34.655037 & 168  & 1.9  \\
07 Jan 2015  &  11:29:05 &  3692.98230  & 35.067655 & 162  & 2.0  \\
07 Jan 2015  &  11:46:50 &  3692.99461  & 35.091275 & 152  & 2.2  \\
07 Jan 2015  &  12:04:35 &  3693.00694  & 35.114932 & 144  & 2.3  \\
08 Jan 2015  &  04:44:36 &  3693.70134  & 36.447285 & 202  & 1.6  \\
08 Jan 2015  &  05:02:30 &  3693.71377  & 36.471134 & 202  & 1.6  \\
08 Jan 2015  &  05:20:14 &  3693.72609  & 36.494773 & 202  & 1.5  \\
08 Jan 2015  &  10:30:48 &  3693.94174  & 36.908543 & 191  & 1.7  \\
08 Jan 2015  &  10:48:33 &  3693.95406  & 36.932181 & 188  & 1.7  \\
08 Jan 2015  &  11:06:18 &  3693.96639  & 36.955839 & 191  & 1.7  \\
09 Jan 2015  &  04:39:52 &  3694.69797  & 38.359529 & 189  & 1.7  \\
09 Jan 2015  &  04:57:37 &  3694.71030  & 38.383187 & 189  & 1.7  \\
09 Jan 2015  &  05:15:22 &  3694.72262  & 38.406825 & 193  & 1.6  \\
09 Jan 2015  &  10:24:48 &  3694.93749  & 38.819099 & 191  & 1.7  \\
09 Jan 2015  &  10:42:34 &  3694.94982  & 38.842756 & 190  & 1.7  \\
09 Jan 2015  &  11:00:19 &  3694.96215  & 38.866414 & 178  & 1.8  \\
10 Jan 2015  &  04:40:42 &  3695.69846  & 40.279180 & 182  & 1.7  \\
10 Jan 2015  &  04:58:27 &  3695.71079  & 40.302838 & 177  & 1.8  \\
10 Jan 2015  &  05:16:11 &  3695.72311  & 40.326476 & 183  & 1.8  \\
10 Jan 2015  &  10:26:58 &  3695.93890  & 40.740515 & 190  & 1.7  \\
10 Jan 2015  &  10:44:43 &  3695.95123  & 40.764172 & 190  & 1.7  \\
10 Jan 2015  &  11:02:29 &  3695.96356  & 40.787830 & 180  & 1.8  \\
12 Jan 2015  &  10:25:28 &  3697.93769  & 44.575614 & 198  & 1.6  \\
12 Jan 2015  &  10:43:13 &  3697.95002  & 44.599271 & 195  & 1.7  \\
		\hline		
	\end{tabular}
	\label{tab:Data}
\end{table*}

\section{Stokes signatures}
\begin{figure*}
    \centering
    \includegraphics[width=1.9\columnwidth]{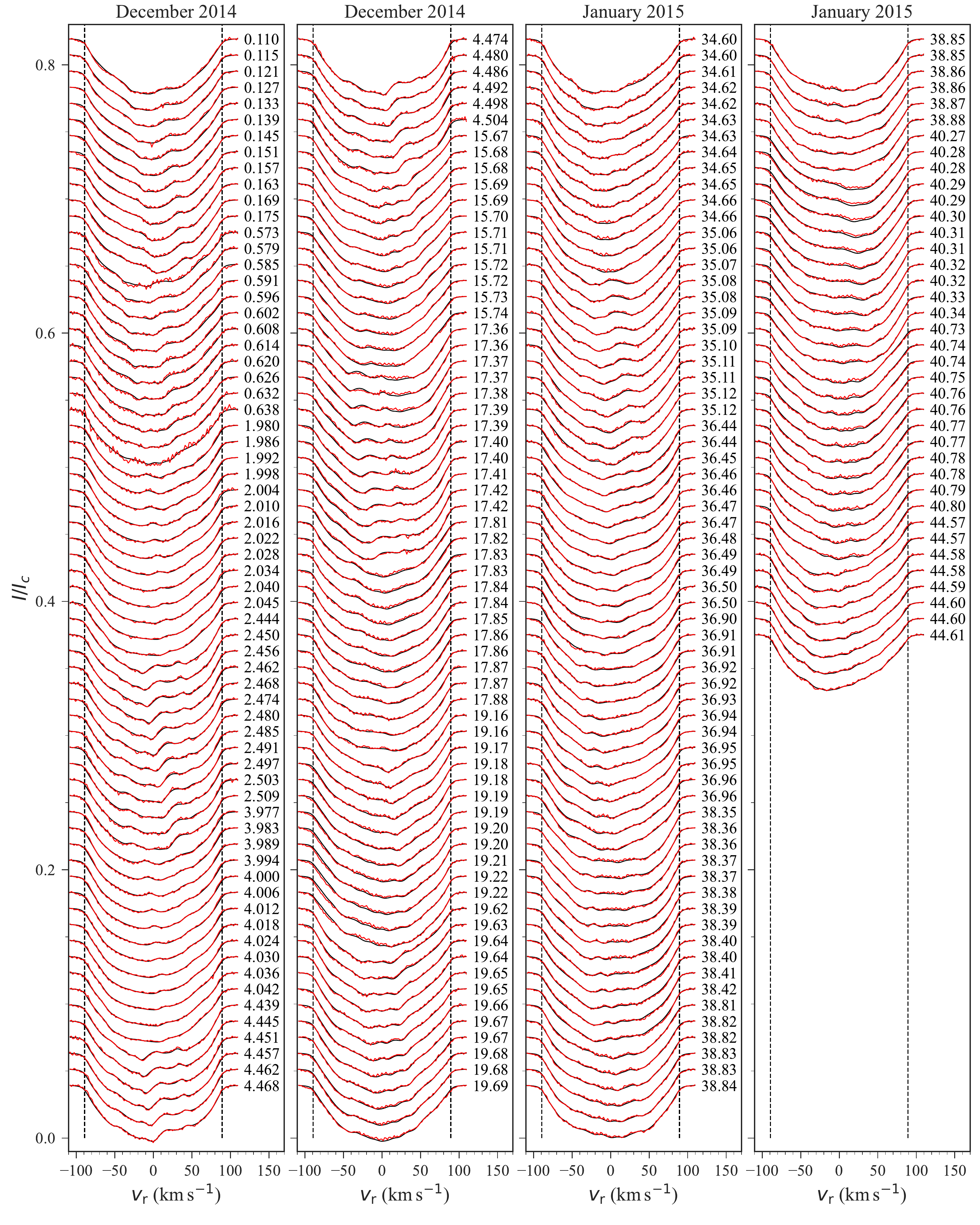}
    \caption{Stokes $I$ profiles in 2014.9 (\BZ{columns 1 and 2}) and 2015.1 (\BZ{columns 3 and 4}) data sets. Observed Stokes $I$ LSD profiles are shown in red, whereas modelled observations are shown in black (see Section~\ref{sec:var} for further details). All profiles are equally shifted for illustration purposes. The rotation cycle of each observation is shown in the right.}
   \label{fig:stokesI}
\end{figure*}

\begin{figure*}
    \centering
    \includegraphics[width=1.5\columnwidth]{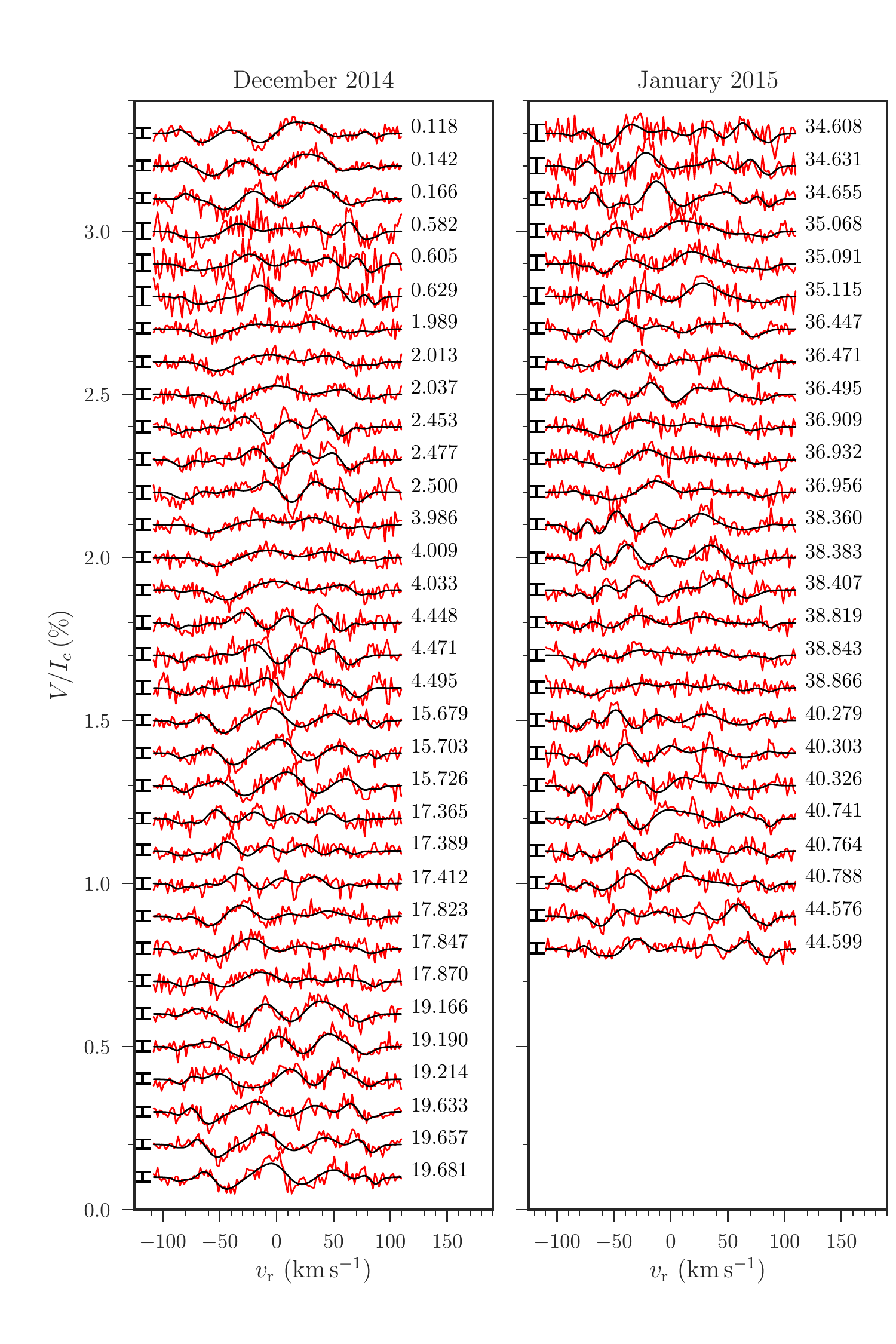}
    \caption{Stokes $V$ profiles in 2014.9 (left) and 2015.1 (right) data sets. Observed Stokes $V$ LSD profiles are shown in red, whereas modelled observations are shown in black (see Section~\ref{sec:var} for further details). All profiles are equally shifted for illustration purposes. The rotation cycle of each observation is shown in the right and 1\,$\sigma$ error bars in the left.}
   \label{fig:stokesV}
\end{figure*}

Stokes $I$ and Stokes $V$ profiles are given in Figures~\ref{fig:stokesI} and \ref{fig:stokesV}, respectively. Observed Stokes LSD profiles are shown in red, and modelled Stokes profiles are given in black. Modelled Stokes signatures are associated with the surface maps obtained through independent \verb|ZDI| reconstructions using either the spectropolarimetric data set of December 2014 (top panels in Figure~\ref{fig:mapsplit}) or January 2015 (bottom panels in Figure~\ref{fig:mapsplit}).

\bsp	
\label{lastpage}
\end{document}